\documentclass[12pt,aps,prd,preprint,nofootinbib,longbibliography,superscriptaddress,eqsecnum,amsmath,amsfonts,amssymb]{revtex4-2}
\usepackage{graphicx} 
\usepackage{physics}
\usepackage{xcolor}
\usepackage{hyperref}
\usepackage{tikz}
\usepackage[margin=1in]{geometry}
\usepackage{csquotes}

\allowdisplaybreaks[1]

\begin{document}

\title{What U Can Do:\\ New Solutions and New Challenges Beyond Leading Order}
\date{\today}

\author{Yi Pang}\email{pangyi1@tju.edu.cn}
\affiliation{Center for Joint Quantum Studies and Department of Physics, School of Science, Tianjin University, Tianjin 300350, China}
\affiliation{Peng Huanwu Center for Fundamental Theory, Hefei, Anhui 230026, China}

\author{Robert J. Saskowski}\email{robert\_saskowski@tju.edu.cn (corresponding author)}
\affiliation{Center for Joint Quantum Studies and Department of Physics, School of Science, Tianjin University, Tianjin 300350, China}

\begin{abstract}
    String theories naturally exhibit dualities that lead to hidden symmetries in the low-energy effective description, which have been used to great effect to generate supergravity solutions. We review recent progress in using hidden symmetries arising from T-duality to generate higher-derivative-corrected solutions, as well as the problems that arise from non-perturbative effects when extending this paradigm to hidden symmetries arising from U-duality.
\end{abstract}

\maketitle
\vspace{1cm}
\noindent\emph{Essay written for the Gravity Research Foundation 2026 Awards for Essays on Gravitation}
\addtocounter{page}{-1}
\thispagestyle{empty}

\newpage

\begin{displayquote}
    \emph{Our future discoveries must be looked for in the sixth place of decimals.}
\end{displayquote}
\begin{flushright}
    --Albert A. Michelson
\end{flushright}

\section{Introduction}
Einstein's theory of general relativity is remarkably successful at describing semiclassical gravitational phenomena, but it cannot be the whole story. In particular, the theory's inherent non-renormalizability indicates that it is not UV-complete. As such, it is best thought of as the lowest order in an effective field theory (EFT) expansion,
\begin{equation}
    \mathcal L=\mathcal L_{2\partial}+\Lambda^{-2}_c\mathcal L_{4\partial}+\Lambda_c^{-4}\mathcal L_{6\partial}+\cdots,
\end{equation}
where $\Lambda_c$ is a large cutoff scale. From this perspective, the two-derivative Lagrangian $\mathcal L_{2\partial}$ contains general relativity coupled to some appropriate choice of matter (the Standard Model, dark matter, inflaton, \emph{etc.}), and the higher-derivative corrections $\mathcal L_{(2+2n)\partial}$ encode the contributions of both UV and quantum effects. Such higher-derivative corrections have had important applications ranging from precision holography~\cite{Blau:1999vz,Bobev:2021qxx} to cosmology~\cite{Maldacena:2011nz,Nojiri:2017ncd}.

One of our best candidates for a theory of quantum gravity, and one of the only known UV-complete theories of gravity, is string theory. This gives a distinguished choice of EFT, where the cutoff $\Lambda_c^{-2}$ is identified with the string scale $\alpha'$. This string EFT then takes the form of supergravity with higher-derivative corrections, which can be obtained either by demanding conformal invariance of the worldsheet sigma model or by computing string amplitudes. In particular, the higher-derivative corrections receive contributions from integrating out massive excitations, both perturbative (strings) and non-perturbative (branes), at both the tree (classical) and loop (quantum) levels.

We are generally interested in finding higher-derivative corrections to black hole solutions. However, the direct solution of the Einstein equations is quite difficult and is often possible only for very restricted classes of solutions. One approach to evade this issue is to use hidden symmetries to generate solutions. The study of hidden symmetries originates with Ehlers' 1959 observation that four-dimensional gravity reduced on a circle has an enhanced $SL(2,\mathbb R)$ symmetry~\cite{Ehlers:1959aug}. However, the origin of such hidden symmetries is often quite opaque. One reliable source is string theories. In particular, strings have an extended dimension that leads to a T-duality that swaps winding and momentum modes and persists as a symmetry of the low-energy supergravity description. Often, this T-duality symmetry extends to a larger U-duality group that also includes non-perturbative S-duality transformations, which exchange the strong- and weak-coupling regimes. Such symmetries can then be used to generate black hole solutions without directly solving the equations of motion, providing a systematic way to construct them.

Hitherto, most work on solution generation using hidden symmetries has focused on two-derivative solutions. In this essay, we instead review recent progress in extending T- and U-duality symmetries to the construction of higher-derivative corrected solutions. The remainder of this essay is organized as follows. In Section~\ref{sec:Tduality}, we review the use of T-duality in solution generation, with an emphasis on recent developments in constructing four-derivative corrections. In Section~\ref{sec:Uduality}, we examine the obstruction to extending these methods from T-duality to U-duality in the presence of higher-derivative terms. We conclude in Section~\ref{sec:summary} with a brief summary.

\section{Generating solutions with T-duality}\label{sec:Tduality}
\begin{figure}
    \centering
    \tikzset{every picture/.style={line width=0.75pt}} 
            
            \begin{tikzpicture}[x=0.75pt,y=0.75pt,yscale=-1,xscale=1,scale=1.95]
            
            \draw   (130.57,140.06) -- (424.44,72.91) .. controls (430.07,71.63) and (438.1,85.79) .. (442.39,104.55) .. controls (446.67,123.31) and (445.59,139.56) .. (439.96,140.84) -- (146.09,207.99) .. controls (140.46,209.27) and (132.42,195.11) .. (128.14,176.35) .. controls (123.85,157.59) and (124.94,141.34) .. (130.57,140.06) .. controls (136.19,138.77) and (144.23,152.94) .. (148.52,171.69) .. controls (152.8,190.45) and (151.71,206.7) .. (146.09,207.99) ;
            \draw [color={rgb, 255:red, 74; green, 144; blue, 226 }  ,draw opacity=1 ]   (190.73,168.66) .. controls (197.63,161.46) and (198.49,168.37) .. (207.97,168.66) ;
            \draw [color={rgb, 255:red, 74; green, 144; blue, 226 }  ,draw opacity=1 ]   (207.97,168.66) .. controls (220.22,172.35) and (213.49,182.18) .. (209.87,186.79) ;
            \draw [color={rgb, 255:red, 74; green, 144; blue, 226 }  ,draw opacity=1 ]   (192.63,186.79) .. controls (199.52,180.51) and (202.97,193.07) .. (209.87,186.79) ;
            \draw [color={rgb, 255:red, 74; green, 144; blue, 226 }  ,draw opacity=1 ]   (181.76,175.9) .. controls (179.52,177.58) and (185.73,193.07) .. (192.63,186.79) ;
            \draw [color={rgb, 255:red, 74; green, 144; blue, 226 }  ,draw opacity=1 ]   (181.76,175.9) .. controls (191.25,170.88) and (188.66,171.51) .. (190.73,168.66) ;
            
            \draw [color={rgb, 255:red, 74; green, 144; blue, 226 }  ,draw opacity=1 ]   (196,156.6) .. controls (190.24,115.32) and (176.19,113.67) .. (159.15,126.87) ;
            \draw [shift={(157,128.6)}, rotate = 320.19] [fill={rgb, 255:red, 74; green, 144; blue, 226 }  ,fill opacity=1 ][line width=0.08]  [draw opacity=0] (8.93,-4.29) -- (0,0) -- (8.93,4.29) -- cycle    ;
            \draw  [draw opacity=0] (315.14,98.16) .. controls (315.14,98.16) and (315.14,98.16) .. (315.14,98.16) .. controls (320.54,96.8) and (328.66,110.5) .. (333.28,128.76) .. controls (337.9,147.02) and (337.26,162.93) .. (331.86,164.3) -- (323.5,131.23) -- cycle ; \draw  [color={rgb, 255:red, 208; green, 2; blue, 27 }  ,draw opacity=1 ] (315.14,98.16) .. controls (315.14,98.16) and (315.14,98.16) .. (315.14,98.16) .. controls (320.54,96.8) and (328.66,110.5) .. (333.28,128.76) .. controls (337.9,147.02) and (337.26,162.93) .. (331.86,164.3) ;  
            \draw  [draw opacity=0] (325.14,96.16) .. controls (325.14,96.16) and (325.14,96.16) .. (325.14,96.16) .. controls (330.54,94.8) and (338.66,108.5) .. (343.28,126.76) .. controls (347.9,145.02) and (347.26,160.93) .. (341.86,162.3) -- (333.5,129.23) -- cycle ; \draw  [color={rgb, 255:red, 208; green, 2; blue, 27 }  ,draw opacity=1 ] (325.14,96.16) .. controls (325.14,96.16) and (325.14,96.16) .. (325.14,96.16) .. controls (330.54,94.8) and (338.66,108.5) .. (343.28,126.76) .. controls (347.9,145.02) and (347.26,160.93) .. (341.86,162.3) ;  
            
            \draw (179,200.4) node [anchor=north west][inner sep=0.75pt]  [color={rgb, 255:red, 74; green, 144; blue, 226 }  ,opacity=1 ]  {$p=\frac{n}{R}$};
            \draw (312,167.4) node [anchor=north west][inner sep=0.75pt]  [color={rgb, 255:red, 208; green, 2; blue, 27 }  ,opacity=1 ]  {$\tilde p=\frac{mR}{\alpha '}$};
            \end{tikzpicture}
    \caption{Illustration of momentum (blue) and winding (red) modes for strings compactified on a circle.}
    \label{fig:Tduality}
\end{figure}
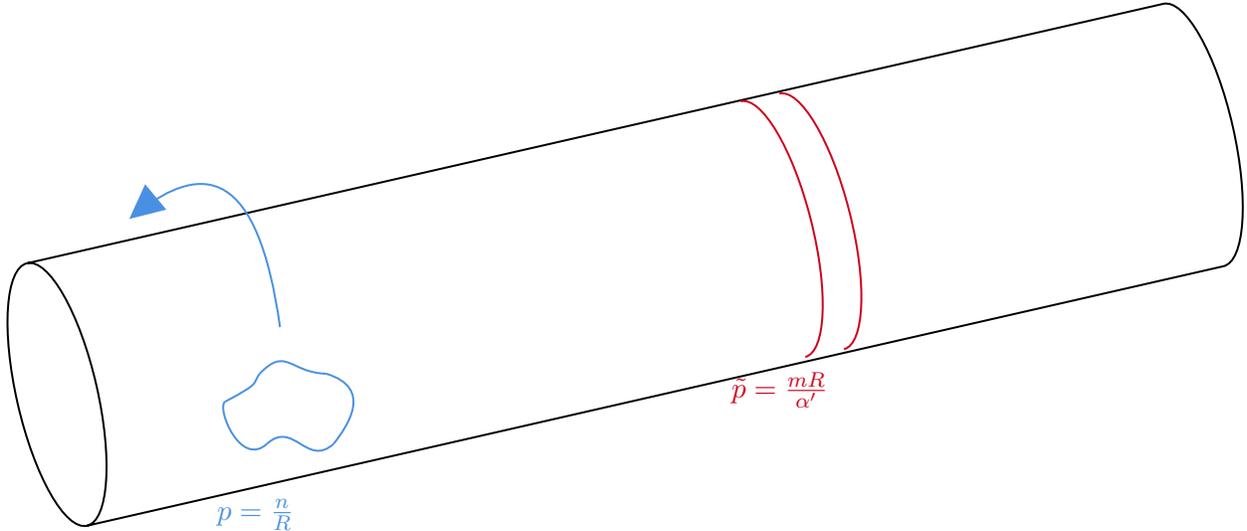

If we put a theory of point particles, such as gravity, on a $d$-dimensional torus $T^d$, then those particles can have momentum along the $d$ cycles of the torus. Since each cycle is compact, these internal momenta will correspond to Fourier modes $p^i=n^i/R_i$ ($i=1,...,d$), where $n^i\in\mathbb Z$ and $R_i$ is the radius of the $i^\mathrm{th}$ circle. However, strings have an additional extended dimension, which can wrap around the cycles of the torus. Thus, the compactified string is also characterized by winding modes $\tilde p_i=w_i R_i/\alpha'$, where the winding number $w_i\in\mathbb Z$ characterizes how many times the string wraps around the $i^\mathrm{th}$ circle, and the factor of $R_i/\alpha'$ comes from the product of the circle circumference ($2\pi R_i$) with the string tension ($1/2\pi\alpha'$). This is depicted in Figure~\ref{fig:Tduality}. Consequently, string theory has a T-duality that swaps the momentum and winding modes $p^i\leftrightarrow\tilde p_i$, which is equivalent to swapping $n^i\leftrightarrow w_i$ and inverting the circle radius $R_i\to\alpha'/R_i$. For the heterotic string, there is a third kind of momentum corresponding to the fact that the left-movers are ten-dimensional supersymmetric modes and the right-movers are 26-dimensional bosonic modes. As such, the extra 16 dimensions of the bosonic modes are reduced on a torus, which leads to 16 internal momentum modes $P^{\mathfrak a}$ ($\mathfrak a=1,...,16$) that can intermix with the momentum and winding modes. 

The most general background in which the string might propagate has a nontrivial metric, Neveu-Schwarz (NS) two-form, and dilaton. Putting this on a toroidal background means that the ten-dimensional spacetime $\mathcal M_{10}$ can be written as a torus $T^d$ fibered over some $(10-d)$-dimensional noncompact spacetime $\mathcal M_{10-d}$. The metric then splits into a $(10-d)$-dimensional metric $g_{\mu\nu}$, a principal connection $A^i_\mu$ for the fibration, and an internal metric $g_{ij}$ on the torus. $A^i_\mu$ then transforms as a gauge field under the isometry group of the torus, $U(1)^d$, while $g_{ij}$ transforms as a matrix of scalars. Similarly, the NS two-form $B_{\mu\nu}$ splits into a $(10-d)$-dimensional two-form $b_{\mu\nu}$, a gauge field $B_{\mu i}$, and an antisymmetric matrix of scalars $b_{ij}$. For the heterotic string, there is also a gauge field transforming under $E_8\times E_8$ or $SO(32)$. To see a further symmetry enhancement, this must be broken by the background to an abelian subgroup, $U(1)^{p}$, with $p\le 16$. As such, the ten-dimensional heterotic gauge fields on a torus will become a $(10-d)$-dimensional gauge field $\mathcal A^{\mathfrak a}_\mu$ ($\mathfrak a=1,...,p$) and a matrix of scalars $\mathcal A_i^{\mathfrak a}$.

Placed on a torus, the string level matching condition, which corresponds to the requirement that a closed string have no distinguished origin, becomes~\cite{Giveon:1994fu}
\begin{equation}
    0=\mathcal P^T\eta\mathcal P+\mathrm{oscillators},
\end{equation}
where $\mathcal P=(n^i, w_i, P^{\mathfrak a})$ is the \emph{generalized momentum} and $\eta$ is a constant matrix. The oscillators correspond to excited string modes, which will not directly affect the low-energy limit we are concerned with. Clearly, this is invariant under the linear transformation $\mathcal P\to\Omega\mathcal P$, so long as it leaves $\eta$ invariant, $\Omega^T\eta\Omega=\eta$. This condition is simply the requirement that $\Omega$ be an $O(d+p,d;\mathbb Z)$ transformation, where the restriction to integer coefficients is because the momentum and winding modes must remain integers. Similarly, the string mass spectrum on a torus takes the form
\begin{equation}
    \alpha'M^2=\mathcal P^T\mathcal H\mathcal P+\mathrm{oscillators},
\end{equation}
where the \emph{generalized metric} $\mathcal H$ is a matrix built out of the scalars $g_{ij}$, $b_{ij}$, and $\mathcal A_i^{\mathfrak a}$. This is invariant under $O(d+p,d;\mathbb Z)$ transformations, as long as $\mathcal H$ transforms as ${\mathcal H\to(\Omega^{-1})^T\mathcal H\Omega^{-1}}$. This extends to a symmetry of the entire mass spectrum.

In the low-energy limit, we neglect the existence of strings and approximate the theory by a supergravity theory describing point particles. But our effective theory does not forget its stringy origins, and the $O(d+p,d;\mathbb Z)$ duality persists as a symmetry of heterotic supergravity. In fact, going to the classical limit, we forget about the quantization of charges, and correspondingly, the discrete duality group $O(d+p,d;\mathbb Z)$ is enhanced to a continuous $O(d+p,d;\mathbb R)$ symmetry. Thus, reducing the two-derivative heterotic supergravity on a torus then automatically leads to a coset model~\cite{Maharana:1992my}
\begin{equation}
    e^{-1}\mathcal L=e^{-2\varphi}\qty[R-4(\partial\varphi)^2-\frac{1}{12}h_{\mu\nu\rho}h^{\mu\nu\rho}+\frac{1}{8}\Tr\qty(\partial_\mu\mathcal H\partial^\mu\mathcal H^{-1})-\frac{1}{4}\mathbb F^T_{\mu\nu}\mathcal H\mathbb F^{\mu\nu}],
\end{equation}
where $\mathbb F=\dd\mathbb A$, and the \emph{generalized gauge field} $\mathbb A$ is a vector built from the gauge fields $A^i_\mu$, $B_{\mu i}$, and $\mathcal A_\mu^{\mathfrak a}$. Here, $h_{\mu\nu\rho}$ is the field strength of $b_{\mu\nu}$ and $\varphi$ is the dilaton. This is invariant under the global $O(d+p,d;\mathbb R)$ transformation
\begin{equation}
    \mathcal H\to\qty(\Omega^{-1})^T\mathcal H\Omega^{-1},\qquad\mathbb A_\mu\to\Omega\mathbb A_\mu,\label{eq:OdpdTrans}
\end{equation}
with all other fields remaining invariant. Such a transformation can be applied anytime that spacetime contains a torus.

Taking a step back, suppose we have a solution to heterotic supergravity with a $U(1)^d$ isometry. Then there is a collection of $d$ abelian Killing vectors $K_i$ ($i=1,...,d$), which automatically defines a set of coordinates $K_i\equiv\partial/\partial y^i$ describing a torus. That is, a heterotic solution with a $U(1)^d$ isometry is equivalent to compactifying on $T^d$, which means that there is automatically an $O(d+p,d;\mathbb R)$ symmetry present. In this case, reducing along this torus, transforming the fields as in~\eqref{eq:OdpdTrans}, and then uplifting will automatically yield a new (inequivalent) solution to the equations of motion. This \emph{Hassan-Sen procedure} was first performed in Ref.~\cite{Hassan:1991mq}.

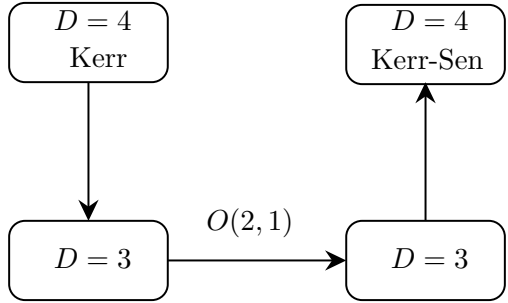
\begin{figure}
    \centering
    \tikzset{every picture/.style={line width=0.75pt}} 
    
    \begin{tikzpicture}[x=0.75pt,y=0.75pt,yscale=-1,xscale=1]
    
    \draw   (110,158) .. controls (110,153.58) and (113.58,150) .. (118,150) -- (182,150) .. controls (186.42,150) and (190,153.58) .. (190,158) -- (190,182) .. controls (190,186.42) and (186.42,190) .. (182,190) -- (118,190) .. controls (113.58,190) and (110,186.42) .. (110,182) -- cycle ;
    \draw   (280,159) .. controls (280,154.58) and (283.58,151) .. (288,151) -- (352,151) .. controls (356.42,151) and (360,154.58) .. (360,159) -- (360,183) .. controls (360,187.42) and (356.42,191) .. (352,191) -- (288,191) .. controls (283.58,191) and (280,187.42) .. (280,183) -- cycle ;
    \draw   (110,268) .. controls (110,263.58) and (113.58,260) .. (118,260) -- (182,260) .. controls (186.42,260) and (190,263.58) .. (190,268) -- (190,292) .. controls (190,296.42) and (186.42,300) .. (182,300) -- (118,300) .. controls (113.58,300) and (110,296.42) .. (110,292) -- cycle ;
    \draw   (280,268) .. controls (280,263.58) and (283.58,260) .. (288,260) -- (352,260) .. controls (356.42,260) and (360,263.58) .. (360,268) -- (360,292) .. controls (360,296.42) and (356.42,300) .. (352,300) -- (288,300) .. controls (283.58,300) and (280,296.42) .. (280,292) -- cycle ;
    \draw    (150,190) -- (150,257) ;
    \draw [shift={(150,260)}, rotate = 270] [fill={rgb, 255:red, 0; green, 0; blue, 0 }  ][line width=0.08]  [draw opacity=0] (10.72,-5.15) -- (0,0) -- (10.72,5.15) -- (7.12,0) -- cycle    ;
    \draw    (190,280) -- (277,280) ;
    \draw [shift={(280,280)}, rotate = 180] [fill={rgb, 255:red, 0; green, 0; blue, 0 }  ][line width=0.08]  [draw opacity=0] (10.72,-5.15) -- (0,0) -- (10.72,5.15) -- (7.12,0) -- cycle    ;
    \draw    (320,260) -- (320,193) ;
    \draw [shift={(320,190)}, rotate = 90] [fill={rgb, 255:red, 0; green, 0; blue, 0 }  ][line width=0.08]  [draw opacity=0] (10.72,-5.15) -- (0,0) -- (10.72,5.15) -- (7.12,0) -- cycle    ;
    
    \draw (131,152.4) node [anchor=north west][inner sep=0.75pt]    {$D=4$};
    \draw (139,171) node [anchor=north west][inner sep=0.75pt]   [align=left] {Kerr};
    \draw (301,152.4) node [anchor=north west][inner sep=0.75pt]    {$D=4$};
    \draw (291,172) node [anchor=north west][inner sep=0.75pt]   [align=left] {Kerr-Sen};
    \draw (131,272.4) node [anchor=north west][inner sep=0.75pt]    {$D=3$};
    \draw (301,272.4) node [anchor=north west][inner sep=0.75pt]    {$D=3$};
    \draw (208,252.4) node [anchor=north west][inner sep=0.75pt]    {$O( 2,1)$};

    \end{tikzpicture}
    \caption{Schematic depiction of the series of uplifts and transformations used to obtain the two-derivative Kerr-Sen solution.}
    \label{fig:KS}
\end{figure}

If we focus on four-dimensional black hole solutions and keep only one of the heterotic gauge fields ($p=1$), then one gets the \emph{Kerr-Sen solution}~\cite{Sen:1991zi}. The key observation is that the Kerr solution is time-independent, which leads to an $\mathbb R$ symmetry, which is the noncompact form of $U(1)$. This, in turn, leads to an $O(2,1)$ symmetry that can be used to generate solutions. Of the three $O(2,1)$ generators, only one of them acts nontrivially on solutions, with the other two corresponding to large diffeomorphisms and gauge transformations. Applying this transformation, depicted in Figure~\ref{fig:KS}, generates a non-extremal, charged, rotating black hole solution that is the most general axisymmetric solution to the heterotic theory with one gauge field~\cite{Rogatko:2010hf}.

We would naturally like to extend this to higher-derivative corrections. Fortunately, the $O(d+p,d;\mathbb R)$ symmetry of heterotic supergravity on $T^d$ persists to all orders in the tree-level $\alpha'$-expansion~\cite{Sen:1991zi,Hohm:2014sxa}. However, there is the subtlety of field redefinition ambiguities. Since $\alpha'$ is treated perturbatively, we can always redefine our fields schematically as ${\Phi\to\Phi+\alpha'\delta\Phi}$, which causes the action to shift
\begin{equation}
    \mathcal L\to\mathcal L+\alpha'\mathcal E_\Phi\delta\Phi+\mathcal O(\alpha'^2),
\end{equation}
where $\mathcal E_\Phi$ are the equations of motion corresponding to the field $\Phi$. Consequently, we can always shift the higher-derivative action by the two-derivative equations of motion. Unlike in the two-derivative case, the $O(d+p,d;\mathbb R)$ symmetry is not manifestly present in the higher-derivative terms in the action and must be restored via appropriate field redefinitions~\cite{Eloy:2020dko,Elgood:2020xwu,Ortin:2020xdm,Jayaprakash:2024xlr}. This is equivalent to an $\alpha'$-deformation of the $O(d+p,d;\mathbb R)$ transformation rules acting on the fields in the original field redefinition frame~\cite{Kaloper:1997ux,Garousi:2019wgz}.

One issue is that many of these higher-derivative results regarding heterotic supergravity truncate the heterotic gauge fields. However, there is a trick to restore them: One can start $p$ dimensions higher, reduce on $T^p$, and then apply a consistent truncation to reinstate the gauge fields. There are two possible consistent choices of consistent truncation, corresponding to $A^i=\pm B_i$. These lead to the same action at the two-derivative level but to two distinct actions at the four-derivative level. One of these corresponds to the usual four-derivative heterotic action with gauge fields~\cite{Xia:2025lvn}, whereas the other corresponds to a non-supersymmetric extension of the heterotic theory~\cite{Hu:2025aji}. Moreover, these are the only two four-derivative actions that have an $O(d+p,d;\mathbb R)$ symmetry upon dimensional reduction on $T^d$~\cite{Baron:2017dvb}.

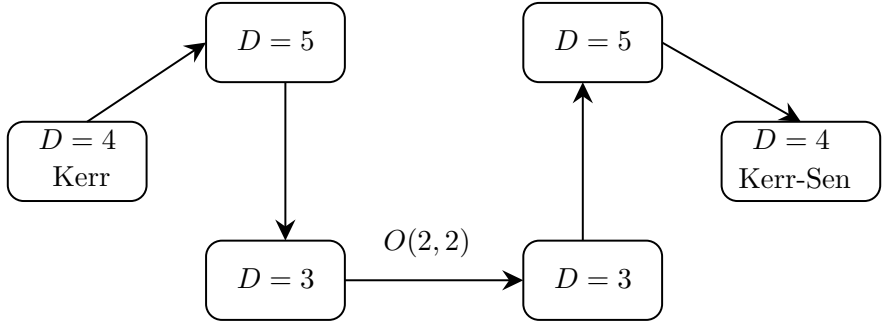
\begin{figure}
    \centering
    \tikzset{every picture/.style={line width=0.75pt}} 

    \begin{tikzpicture}[x=0.75pt,y=0.75pt,yscale=-1,xscale=1]
    
    \draw   (40,118) .. controls (40,113.58) and (43.58,110) .. (48,110) -- (102,110) .. controls (106.42,110) and (110,113.58) .. (110,118) -- (110,142) .. controls (110,146.42) and (106.42,150) .. (102,150) -- (48,150) .. controls (43.58,150) and (40,146.42) .. (40,142) -- cycle ;
    \draw   (140,58) .. controls (140,53.58) and (143.58,50) .. (148,50) -- (202,50) .. controls (206.42,50) and (210,53.58) .. (210,58) -- (210,82) .. controls (210,86.42) and (206.42,90) .. (202,90) -- (148,90) .. controls (143.58,90) and (140,86.42) .. (140,82) -- cycle ;
    \draw   (140,178) .. controls (140,173.58) and (143.58,170) .. (148,170) -- (202,170) .. controls (206.42,170) and (210,173.58) .. (210,178) -- (210,202) .. controls (210,206.42) and (206.42,210) .. (202,210) -- (148,210) .. controls (143.58,210) and (140,206.42) .. (140,202) -- cycle ;
    \draw   (300,58) .. controls (300,53.58) and (303.58,50) .. (308,50) -- (362,50) .. controls (366.42,50) and (370,53.58) .. (370,58) -- (370,82) .. controls (370,86.42) and (366.42,90) .. (362,90) -- (308,90) .. controls (303.58,90) and (300,86.42) .. (300,82) -- cycle ;
    \draw   (300,178) .. controls (300,173.58) and (303.58,170) .. (308,170) -- (362,170) .. controls (366.42,170) and (370,173.58) .. (370,178) -- (370,202) .. controls (370,206.42) and (366.42,210) .. (362,210) -- (308,210) .. controls (303.58,210) and (300,206.42) .. (300,202) -- cycle ;
    \draw   (400,118) .. controls (400,113.58) and (403.58,110) .. (408,110) -- (472,110) .. controls (476.42,110) and (480,113.58) .. (480,118) -- (480,142) .. controls (480,146.42) and (476.42,150) .. (472,150) -- (408,150) .. controls (403.58,150) and (400,146.42) .. (400,142) -- cycle ;
    \draw [color={rgb, 255:red, 0; green, 0; blue, 0 }  ,draw opacity=1 ]   (80,110) -- (137.5,71.66) ;
    \draw [shift={(140,70)}, rotate = 146.31] [fill={rgb, 255:red, 0; green, 0; blue, 0 }  ,fill opacity=1 ][line width=0.08]  [draw opacity=0] (10.72,-5.15) -- (0,0) -- (10.72,5.15) -- (7.12,0) -- cycle    ;
    \draw [color={rgb, 255:red, 0; green, 0; blue, 0 }  ,draw opacity=1 ]   (180,90) -- (180,167) ;
    \draw [shift={(180,170)}, rotate = 270] [fill={rgb, 255:red, 0; green, 0; blue, 0 }  ,fill opacity=1 ][line width=0.08]  [draw opacity=0] (10.72,-5.15) -- (0,0) -- (10.72,5.15) -- (7.12,0) -- cycle    ;
    \draw [color={rgb, 255:red, 0; green, 0; blue, 0 }  ,draw opacity=1 ]   (210,190) -- (297,190) ;
    \draw [shift={(300,190)}, rotate = 180] [fill={rgb, 255:red, 0; green, 0; blue, 0 }  ,fill opacity=1 ][line width=0.08]  [draw opacity=0] (10.72,-5.15) -- (0,0) -- (10.72,5.15) -- (7.12,0) -- cycle    ;
    \draw [color={rgb, 255:red, 0; green, 0; blue, 0 }  ,draw opacity=1 ]   (330,170) -- (330,93) ;
    \draw [shift={(330,90)}, rotate = 90] [fill={rgb, 255:red, 0; green, 0; blue, 0 }  ,fill opacity=1 ][line width=0.08]  [draw opacity=0] (10.72,-5.15) -- (0,0) -- (10.72,5.15) -- (7.12,0) -- cycle    ;
    \draw [color={rgb, 255:red, 0; green, 0; blue, 0 }  ,draw opacity=1 ]   (370,70) -- (437.4,108.51) ;
    \draw [shift={(440,110)}, rotate = 209.74] [fill={rgb, 255:red, 0; green, 0; blue, 0 }  ,fill opacity=1 ][line width=0.08]  [draw opacity=0] (10.72,-5.15) -- (0,0) -- (10.72,5.15) -- (7.12,0) -- cycle    ;
    
    \draw (61,131) node [anchor=north west][inner sep=0.75pt]   [align=left] {Kerr};
    \draw (54,112.4) node [anchor=north west][inner sep=0.75pt]    {$D=4$};
    \draw (154,62.4) node [anchor=north west][inner sep=0.75pt]    {$D=5$};
    \draw (154,182.4) node [anchor=north west][inner sep=0.75pt]    {$D=3$};
    \draw (314,62.4) node [anchor=north west][inner sep=0.75pt]    {$D=5$};
    \draw (314,182.4) node [anchor=north west][inner sep=0.75pt]    {$D=3$};
    \draw (414,112.4) node [anchor=north west][inner sep=0.75pt]    {$D=4$};
    \draw (407,132) node [anchor=north west][inner sep=0.75pt]   [align=left] {Kerr-Sen};
    \draw (228,162.4) node [anchor=north west][inner sep=0.75pt]    {$O( 2,2)$};

    \end{tikzpicture}
    \caption{Schematic depiction of the series of uplifts, field redefinitions, and transformations we perform to obtain the four-derivative Kerr-Sen solution.}
    \label{fig:KS4der}
\end{figure}

To extend the solution generation procedure to generate the higher-derivative Kerr-Sen solution, one must uplift to five dimensions to restore the gauge fields, reduce to three dimensions, perform the field redefinition to go to the $O(2,2;\mathbb R)$ covariant frame, apply the $O(2,1;\mathbb R)\subset O(2,2;\mathbb R)$ transformation, and then field redefine back to the original field redefinition frame, uplift to five dimensions, and finally reduce back down to four dimensions. This is depicted schematically in Figure~\ref{fig:KS4der}. 
The detour through five dimensions implements the trick for restoring the heterotic gauge fields. Although technically complex, the process itself is straightforward. This extension to higher derivatives has recently been accomplished in the case of four-derivative corrections to Kerr-Sen black holes for both possible four-derivative actions~\cite{Hu:2025aji,Xia:2025lvn}. Interestingly, although the Kerr-Sen and Kerr-Newman black holes have the same gravitational and electromagnetic multipole moments at the two-derivative level, they are distinct at the four-derivative level, indicating that high-precision gravitational-wave experiments could, in principle, distinguish them. This opens the door to using this technique to compute four-derivative corrections to any two-derivative solution obtained using the original Hassan-Sen procedure in heterotic supergravity, such as~\cite{Veneziano:1991ek,Meissner:1991zj,Sen:1991zi,Sen:1991cn,Gasperini:1991qy,Hassan:1991mq,Sen:1992ua,Cvetic:1995sz,Cvetic:1995kv,Cvetic:1996xz}.

\section{(Not) generating solutions with U-duality}\label{sec:Uduality}
In addition to perturbative fundamental strings, string theory also contains branes, which are non-perturbative solitonic objects. A $p$-form gauge field naturally couples to a $p$-dimensional worldvolume, so in this sense, one may think of the string as being the charged object sourcing the NS two-form field $B_{(2)}$. The magnetic dual is a six-form field that is sourced by five-dimensional NS5-branes. Likewise, the $(p+1)$-form Ramond-Ramond (RR) fields $C_{(p+1)}$ of type II string theories are sourced by $p$-dimensional D$p$-branes.

Type IIB string theory has RR fields $C_{(0)}$, $C_{(2)}$, $C_{(4)}$ (and their magnetic duals $C_{(8)}$, $C_{(6)}$, $C_{(4)}$, respectively), which correspond to D$(-1)$, D1, and D3 branes (and their magnetic duals D7, D5, and D3 branes, respectively). As such, there is an S-duality symmetry that exchanges strings with D1-branes and NS5-branes with D5-branes. As the four-form $C_{(4)}$ is self-dual, the D3-brane is its own magnetic dual. This amounts to an inversion of the string coupling $g_s\to1/g_s$. However, the S-duality group is actually $SL(2,\mathbb Z)$. In particular, $C_{(0)}$ combines with the dilaton into a complex scalar that transforms under $SL(2,\mathbb Z)$ via modular transformations, and $C_{(2)}$ and $B_{(2)}$ transform as an $SL(2,\mathbb Z)$ doublet. When IIB strings are placed on a torus $T^{d}$, we get both T- and S-dualities. As these do not commute, they are completed by a larger group of U-dualities, corresponding to $E_{d+1(d+1)}(\mathbb Z)$, which is the noncompact form of the (discrete) exceptional group $E_{d+1}(\mathbb Z)$. A special case is $d=0$, which corresponds to the $E_{1(1)}(\mathbb Z)\cong SL(2,\mathbb Z)$ S-duality of the uncompactified IIB theory. However, the group $E_{d+1(d+1)}(\mathbb Z)$ is far larger than what we need just to encompass $SL(2,\mathbb Z)$ and $O(d,d;\mathbb Z)$. When we reduce on $T^d$, D$p$-branes of dimension $p\le d-1$ can wrap around the compact $(p+1)$-cycles of the torus. This then leads to momentum and winding modes, much as we saw for the fundamental string in T-duality. There is then a duality that exchanges the momentum and winding quantum numbers of these wrapped D-brane instantons. 

In contrast, the heterotic string has no D-branes, but it does have NS5-branes. Reduced on $T^4$, an NS5-brane wrapping the torus will appear as a string-like object~\cite{Font:1990gx}.\footnote{Such NS5-instantons are also involved in the enhancement of the IIB U-duality group.} There is then a symmetry that exchanges the fundamental string and the NS5 ``string.'' Further reduced on $T^2$, we will get an $O(2,2;\mathbb Z)_T$ from T-duality of the string and an $O(2,2;\mathbb Z)_S$ from ``T-duality'' of the NS5-string, which exchanges momentum and winding modes of NS5-branes wrapping five-cycles of $T^6$. These groups can be broken up as $O(2,2;\mathbb Z)_T\cong SL(2,\mathbb Z)_O\times SL(2,\mathbb Z)_T$ and $O(2,2;\mathbb Z)_S\cong SL(2,\mathbb Z)_O\times SL(2,\mathbb Z)_S$, where $SL(2,\mathbb Z)_O$ is a common subgroup corresponding to large diffeomorphisms and gauge transformations~\cite{Duff:1994zt}. Thus, a nontrivial $SL(2,\mathbb Z)$ S-duality arises in four dimensions. We can further reduce to three dimensions: Since the heterotic theory is now reduced on $T^7$, there are seven circles we can identify with the original uncompactified fourth dimension. This leads to seven copies of the $SL(2,\mathbb Z)$ S-duality group, in addition to the $O(23,7;\mathbb Z)$ T-duality group~\cite{Sen:1994fa}. Together, these generate a symmetry enhancement to the $O(24,8;\mathbb Z)$ U-duality group.

In the low-energy effective description, U-duality becomes a continuous symmetry of the theory. In the four-dimensional heterotic case, the NS two-form $b_{(2)}$ can be dualized to a scalar $\tilde b_{(0)}$ which combines with the dilaton into a complex scalar that transforms under modular transformations. Likewise, the generalized gauge field $\mathbb A_{(1)}$ and its dual $\tilde{\mathbb A}_{(1)}$ transform as an $SL(2,\mathbb R)$ doublet. In three dimensions, all the vector fields can be dualized to scalars, and all of the scalars can be packaged inside one large $O(24,8;\mathbb R)$ generalized metric $\mathcal M$, such that the theory becomes a coset model in the Einstein frame~\cite{Sen:1994wr},
\begin{equation}
    e^{-1}\mathcal L=R+\frac{1}{8}\Tr\qty(\partial_\mu\mathcal M\partial^\mu\mathcal M),
\end{equation}
invariant under the $O(24,8;\mathbb R)$ transformation $\mathcal M\to(\Omega^{-1})^T\mathcal M\Omega^{-1}$. There is a similar three-dimensional $E_{8(8)}(\mathbb R)$ coset model for IIB supergravity on $T^7$~\cite{Cremmer:1997ct,Cremmer:1998px}. Likewise, minimal five-dimensional supergravity, which is not directly the low-energy effective description of a string theory, exhibits a $G_{2(2)}(\mathbb R)$ symmetry when reduced on $T^2$~\cite{Cecotti:1988qn,Mizoguchi:1998wv,Cremmer:1999du,Cremmer:1997ct,Cremmer:1998px}, where $G_{2(2)}(\mathbb R)$ is the noncompact form of the exceptional Lie group $G_2(\mathbb R)$. As minimal supergravity can be obtained as a consistent truncation of heterotic or IIB supergravity, this hidden symmetry enhancement to $G_{2(2)}(\mathbb R)$ is interpreted as coming from U-duality. Such coset models can then be used for generating solutions, much as their T-duality counterparts were. This has been put to especially good use for the $G_{2(2)}(\mathbb R)$~\cite{Bouchareb:2007ax,Tomizawa:2008qr,Compere:2009zh,Suzuki:2024coe,Suzuki:2024phv,Suzuki:2024vzq} and $O(4,4;\mathbb R)$~\cite{Chong:2004na,Galtsov:2008bmt,Galtsov:2008jjb}\footnote{Note that the $O(4,4;\mathbb R)$ case is for the five-dimensional STU model reduced on $T^2$. This can be obtained by reducing six-dimensional heterotic supergravity on a circle and truncating the heterotic gauge fields. As such, this corresponds to a $O(4,4;\mathbb R)\subset O(24,8;\mathbb R)$ subgroup.} symmetries, although it remains largely unexploited for larger U-duality groups, such as $E_{d+1(d+1)}(\mathbb R)$.

Naturally, we would like to extend these symmetries to generating higher-derivative solutions. However, the symmetry enhancement to the U-duality group generally involves a classical scaling transformation that scales the action by an overall factor while leaving the equations of motion invariant. In $D$ dimensions, the $n$-derivative Lagrangian is a dimension $D+n-2$ operator. Consequently, a classical scaling transformation generally causes the higher-derivative terms to scale differently from the two-derivative Lagrangian~\cite{Lambert:2006he,Bao:2007er}. As a result, the U-duality group is reduced to a subgroup. Moreover, it has been shown that the loss of this scaling symmetry is enough to eliminate all U-duality enhancement in the case of the heterotic $O(24,8;\mathbb R)$~\cite{Eloy:2022vsq,Pang:2026urr} and the minimal supergravity $G_{2(2)}(\mathbb R)$~\cite{Pang:2026urr} in three dimensions. One might hope that restricting to the discrete symmetry groups could remedy this issue. However, $O(24,8;\mathbb Z)$ and $G_{2(2)}(\mathbb Z)$ are still fully broken to their geometric subgroups by the higher-derivative corrections. Likewise, there is evidence that $E_{8(8)}(\mathbb Z)$ must also be at least partially broken~\cite{Lambert:2006he}.

The issue seems to be the non-perturbative nature of S-duality. After all, the duality exchanges perturbative string states with non-perturbative brane states. This issue is already evident in the IIB theory. There, the leading higher-derivative corrections appear at eight derivatives. The form of the eight-derivative action is fixed by supersymmetry, up to an overall scalar prefactor. The tree-level (Einstein frame) contributions lead to a dilaton prefactor of $e^{-3\phi/2}$, which is not $SL(2,\mathbb Z)$ invariant as it does not involve the RR zero-form $C_{(0)}$. Similarly, one-loop contributions contribute $e^{-\phi/2}$, and higher loops do not contribute at all. But if one takes into account the contribution of D$(-1)$-branes, then the scalar prefactors sum together to become an $SL(2,\mathbb Z)$-invariant non-holomorphic Eisenstein series. This is straightforward, as we can compute the corrections from D-instantons in ten dimensions~\cite{Sen:2020cef}.

The NS5-instantons of the heterotic theory do not arise in ten dimensions but only emerge after compactification on a torus, introducing an inherent background dependence. Because the low-energy description ignores heavy solitonic states, their contributions do not appear when we dimensionally reduce, and so the S-duality symmetry that swaps perturbative states with non-perturbative ones does not manifest. Recovering this symmetry requires retaining these solitonic degrees of freedom, effectively necessitating the full string-theoretic framework.  These instanton contributions enter at higher orders in $\alpha'$ and so do not affect the two-derivative effective action, which is why no such problem occurs there. This means that the two-derivative action receives contributions only from the massless states. The tension arises only once higher-derivative corrections are included, where neglecting these non-perturbative effects is no longer consistent with the desired symmetry. In this way, the breakdown of duality at higher orders reflects not a failure of the symmetry itself, but the tension with the effective description.

\section{Summary}\label{sec:summary}
In this essay, we have examined the role of duality symmetries in generating higher-derivative corrections to gravitational solutions. In particular, T-duality is a perturbative symmetry of classical string theory and is therefore realized order-by-order in the tree-level $\alpha'$-expansion of the effective action. This allows solution-generating techniques, in principle, to be extended to arbitrarily high orders in higher derivatives. Building on initial progress by the present authors, this approach offers a promising avenue for constructing new higher-derivative solutions and exploring the structure of string-corrected geometries.

On the other hand, S-duality mixes perturbative and non-perturbative states. This leads to the U-duality symmetry enhancement not appearing in dimensional reductions of higher-derivative supergravity. 
It remains unclear whether there is a way to evade this issue without performing the calculation using the full string-theoretic description. This difficulty is closely tied to the limited understanding of wrapped-brane instanton contributions, aside from the case of D$(-1)$-branes in type IIB theory. Further progress on these questions would help clarify the role of duality symmetries beyond the perturbative regime.

\begin{acknowledgments}
    This work is supported by the National Natural Science Foundation of China (NSFC) under Grants No. 12175164 and No. 12247103.
\end{acknowledgments}

\let\oldaddcontentsline\addcontentsline
\renewcommand{\addcontentsline}[3]{}
\bibliography{cite}

\begin{thebibliography}{51}%
\makeatletter
\providecommand \@ifxundefined [1]{%
 \@ifx{#1\undefined}
}%
\providecommand \@ifnum [1]{%
 \ifnum #1\expandafter \@firstoftwo
 \else \expandafter \@secondoftwo
 \fi
}%
\providecommand \@ifx [1]{%
 \ifx #1\expandafter \@firstoftwo
 \else \expandafter \@secondoftwo
 \fi
}%
\providecommand \natexlab [1]{#1}%
\providecommand \enquote  [1]{``#1''}%
\providecommand \bibnamefont  [1]{#1}%
\providecommand \bibfnamefont [1]{#1}%
\providecommand \citenamefont [1]{#1}%
\providecommand \href@noop [0]{\@secondoftwo}%
\providecommand \href [0]{\begingroup \@sanitize@url \@href}%
\providecommand \@href[1]{\@@startlink{#1}\@@href}%
\providecommand \@@href[1]{\endgroup#1\@@endlink}%
\providecommand \@sanitize@url [0]{\catcode `\\12\catcode `\$12\catcode `\&12\catcode `\#12\catcode `\^12\catcode `\_12\catcode `\%12\relax}%
\providecommand \@@startlink[1]{}%
\providecommand \@@endlink[0]{}%
\providecommand \url  [0]{\begingroup\@sanitize@url \@url }%
\providecommand \@url [1]{\endgroup\@href {#1}{\urlprefix }}%
\providecommand \urlprefix  [0]{URL }%
\providecommand \Eprint [0]{\href }%
\providecommand \doibase [0]{https://doi.org/}%
\providecommand \selectlanguage [0]{\@gobble}%
\providecommand \bibinfo  [0]{\@secondoftwo}%
\providecommand \bibfield  [0]{\@secondoftwo}%
\providecommand \translation [1]{[#1]}%
\providecommand \BibitemOpen [0]{}%
\providecommand \bibitemStop [0]{}%
\providecommand \bibitemNoStop [0]{.\EOS\space}%
\providecommand \EOS [0]{\spacefactor3000\relax}%
\providecommand \BibitemShut  [1]{\csname bibitem#1\endcsname}%
\let\auto@bib@innerbib\@empty
\bibitem [{\citenamefont {Blau}\ \emph {et~al.}(1999)\citenamefont {Blau}, \citenamefont {Narain},\ and\ \citenamefont {Gava}}]{Blau:1999vz}%
  \BibitemOpen
  \bibfield  {author} {\bibinfo {author} {\bibfnamefont {M.}~\bibnamefont {Blau}}, \bibinfo {author} {\bibfnamefont {K.~S.}\ \bibnamefont {Narain}},\ and\ \bibinfo {author} {\bibfnamefont {E.}~\bibnamefont {Gava}},\ }\bibfield  {title} {\bibinfo {title} {{On subleading contributions to the AdS / CFT trace anomaly}},\ }\href {https://doi.org/10.1088/1126-6708/1999/09/018} {\bibfield  {journal} {\bibinfo  {journal} {JHEP}\ }\textbf {\bibinfo {volume} {09}},\ \bibinfo {pages} {018}},\ \Eprint {https://arxiv.org/abs/hep-th/9904179} {arXiv:hep-th/9904179} \BibitemShut {NoStop}%
\bibitem [{\citenamefont {Bobev}\ \emph {et~al.}(2022)\citenamefont {Bobev}, \citenamefont {Hristov},\ and\ \citenamefont {Reys}}]{Bobev:2021qxx}%
  \BibitemOpen
  \bibfield  {author} {\bibinfo {author} {\bibfnamefont {N.}~\bibnamefont {Bobev}}, \bibinfo {author} {\bibfnamefont {K.}~\bibnamefont {Hristov}},\ and\ \bibinfo {author} {\bibfnamefont {V.}~\bibnamefont {Reys}},\ }\bibfield  {title} {\bibinfo {title} {{AdS$_{5}$ holography and higher-derivative supergravity}},\ }\href {https://doi.org/10.1007/JHEP04(2022)088} {\bibfield  {journal} {\bibinfo  {journal} {JHEP}\ }\textbf {\bibinfo {volume} {04}},\ \bibinfo {pages} {088}},\ \Eprint {https://arxiv.org/abs/2112.06961} {arXiv:2112.06961 [hep-th]} \BibitemShut {NoStop}%
\bibitem [{\citenamefont {Maldacena}\ and\ \citenamefont {Pimentel}(2011)}]{Maldacena:2011nz}%
  \BibitemOpen
  \bibfield  {author} {\bibinfo {author} {\bibfnamefont {J.~M.}\ \bibnamefont {Maldacena}}\ and\ \bibinfo {author} {\bibfnamefont {G.~L.}\ \bibnamefont {Pimentel}},\ }\bibfield  {title} {\bibinfo {title} {{On graviton non-Gaussianities during inflation}},\ }\href {https://doi.org/10.1007/JHEP09(2011)045} {\bibfield  {journal} {\bibinfo  {journal} {JHEP}\ }\textbf {\bibinfo {volume} {09}},\ \bibinfo {pages} {045}},\ \Eprint {https://arxiv.org/abs/1104.2846} {arXiv:1104.2846 [hep-th]} \BibitemShut {NoStop}%
\bibitem [{\citenamefont {Nojiri}\ \emph {et~al.}(2017)\citenamefont {Nojiri}, \citenamefont {Odintsov},\ and\ \citenamefont {Oikonomou}}]{Nojiri:2017ncd}%
  \BibitemOpen
  \bibfield  {author} {\bibinfo {author} {\bibfnamefont {S.}~\bibnamefont {Nojiri}}, \bibinfo {author} {\bibfnamefont {S.~D.}\ \bibnamefont {Odintsov}},\ and\ \bibinfo {author} {\bibfnamefont {V.~K.}\ \bibnamefont {Oikonomou}},\ }\bibfield  {title} {\bibinfo {title} {{Modified Gravity Theories on a Nutshell: Inflation, Bounce and Late-time Evolution}},\ }\href {https://doi.org/10.1016/j.physrep.2017.06.001} {\bibfield  {journal} {\bibinfo  {journal} {Phys. Rept.}\ }\textbf {\bibinfo {volume} {692}},\ \bibinfo {pages} {1} (\bibinfo {year} {2017})},\ \Eprint {https://arxiv.org/abs/1705.11098} {arXiv:1705.11098 [gr-qc]} \BibitemShut {NoStop}%
\bibitem [{\citenamefont {Ehlers}(1962)}]{Ehlers:1959aug}%
  \BibitemOpen
  \bibfield  {author} {\bibinfo {author} {\bibfnamefont {J.}~\bibnamefont {Ehlers}},\ }\bibfield  {title} {\bibinfo {title} {{Transformations of static exterior solutions of Einstein's gravitational field equations into different solutions by means of conformal mapping}},\ }\href@noop {} {\bibfield  {journal} {\bibinfo  {journal} {Colloq. Int. CNRS}\ }\textbf {\bibinfo {volume} {91}},\ \bibinfo {pages} {275} (\bibinfo {year} {1962})}\BibitemShut {NoStop}%
\bibitem [{\citenamefont {Giveon}\ \emph {et~al.}(1994)\citenamefont {Giveon}, \citenamefont {Porrati},\ and\ \citenamefont {Rabinovici}}]{Giveon:1994fu}%
  \BibitemOpen
  \bibfield  {author} {\bibinfo {author} {\bibfnamefont {A.}~\bibnamefont {Giveon}}, \bibinfo {author} {\bibfnamefont {M.}~\bibnamefont {Porrati}},\ and\ \bibinfo {author} {\bibfnamefont {E.}~\bibnamefont {Rabinovici}},\ }\bibfield  {title} {\bibinfo {title} {{Target space duality in string theory}},\ }\href {https://doi.org/10.1016/0370-1573(94)90070-1} {\bibfield  {journal} {\bibinfo  {journal} {Phys. Rept.}\ }\textbf {\bibinfo {volume} {244}},\ \bibinfo {pages} {77} (\bibinfo {year} {1994})},\ \Eprint {https://arxiv.org/abs/hep-th/9401139} {arXiv:hep-th/9401139} \BibitemShut {NoStop}%
\bibitem [{\citenamefont {Maharana}\ and\ \citenamefont {Schwarz}(1993)}]{Maharana:1992my}%
  \BibitemOpen
  \bibfield  {author} {\bibinfo {author} {\bibfnamefont {J.}~\bibnamefont {Maharana}}\ and\ \bibinfo {author} {\bibfnamefont {J.~H.}\ \bibnamefont {Schwarz}},\ }\bibfield  {title} {\bibinfo {title} {{Noncompact symmetries in string theory}},\ }\href {https://doi.org/10.1016/0550-3213(93)90387-5} {\bibfield  {journal} {\bibinfo  {journal} {Nucl. Phys. B}\ }\textbf {\bibinfo {volume} {390}},\ \bibinfo {pages} {3} (\bibinfo {year} {1993})},\ \Eprint {https://arxiv.org/abs/hep-th/9207016} {arXiv:hep-th/9207016} \BibitemShut {NoStop}%
\bibitem [{\citenamefont {Hassan}\ and\ \citenamefont {Sen}(1992)}]{Hassan:1991mq}%
  \BibitemOpen
  \bibfield  {author} {\bibinfo {author} {\bibfnamefont {S.~F.}\ \bibnamefont {Hassan}}\ and\ \bibinfo {author} {\bibfnamefont {A.}~\bibnamefont {Sen}},\ }\bibfield  {title} {\bibinfo {title} {{Twisting classical solutions in heterotic string theory}},\ }\href {https://doi.org/10.1016/0550-3213(92)90336-A} {\bibfield  {journal} {\bibinfo  {journal} {Nucl. Phys. B}\ }\textbf {\bibinfo {volume} {375}},\ \bibinfo {pages} {103} (\bibinfo {year} {1992})},\ \Eprint {https://arxiv.org/abs/hep-th/9109038} {arXiv:hep-th/9109038} \BibitemShut {NoStop}%
\bibitem [{\citenamefont {Sen}(1991)}]{Sen:1991zi}%
  \BibitemOpen
  \bibfield  {author} {\bibinfo {author} {\bibfnamefont {A.}~\bibnamefont {Sen}},\ }\bibfield  {title} {\bibinfo {title} {{O(d) x O(d) symmetry of the space of cosmological solutions in string theory, scale factor duality and two-dimensional black holes}},\ }\href {https://doi.org/10.1016/0370-2693(91)90090-D} {\bibfield  {journal} {\bibinfo  {journal} {Phys. Lett. B}\ }\textbf {\bibinfo {volume} {271}},\ \bibinfo {pages} {295} (\bibinfo {year} {1991})}\BibitemShut {NoStop}%
\bibitem [{\citenamefont {Rogatko}(2010)}]{Rogatko:2010hf}%
  \BibitemOpen
  \bibfield  {author} {\bibinfo {author} {\bibfnamefont {M.}~\bibnamefont {Rogatko}},\ }\bibfield  {title} {\bibinfo {title} {{Uniqueness Theorem for Stationary Axisymmetric Black Holes in Einstein-Maxwell-axion-dilaton Gravity}},\ }\href {https://doi.org/10.1103/PhysRevD.82.044017} {\bibfield  {journal} {\bibinfo  {journal} {Phys. Rev. D}\ }\textbf {\bibinfo {volume} {82}},\ \bibinfo {pages} {044017} (\bibinfo {year} {2010})},\ \Eprint {https://arxiv.org/abs/1007.4374} {arXiv:1007.4374 [hep-th]} \BibitemShut {NoStop}%
\bibitem [{\citenamefont {Hohm}\ \emph {et~al.}(2015)\citenamefont {Hohm}, \citenamefont {Sen},\ and\ \citenamefont {Zwiebach}}]{Hohm:2014sxa}%
  \BibitemOpen
  \bibfield  {author} {\bibinfo {author} {\bibfnamefont {O.}~\bibnamefont {Hohm}}, \bibinfo {author} {\bibfnamefont {A.}~\bibnamefont {Sen}},\ and\ \bibinfo {author} {\bibfnamefont {B.}~\bibnamefont {Zwiebach}},\ }\bibfield  {title} {\bibinfo {title} {{Heterotic Effective Action and Duality Symmetries Revisited}},\ }\href {https://doi.org/10.1007/JHEP02(2015)079} {\bibfield  {journal} {\bibinfo  {journal} {JHEP}\ }\textbf {\bibinfo {volume} {02}},\ \bibinfo {pages} {079}},\ \Eprint {https://arxiv.org/abs/1411.5696} {arXiv:1411.5696 [hep-th]} \BibitemShut {NoStop}%
\bibitem [{\citenamefont {Eloy}\ \emph {et~al.}(2020)\citenamefont {Eloy}, \citenamefont {Hohm},\ and\ \citenamefont {Samtleben}}]{Eloy:2020dko}%
  \BibitemOpen
  \bibfield  {author} {\bibinfo {author} {\bibfnamefont {C.}~\bibnamefont {Eloy}}, \bibinfo {author} {\bibfnamefont {O.}~\bibnamefont {Hohm}},\ and\ \bibinfo {author} {\bibfnamefont {H.}~\bibnamefont {Samtleben}},\ }\bibfield  {title} {\bibinfo {title} {{Duality Invariance and Higher Derivatives}},\ }\href {https://doi.org/10.1103/PhysRevD.101.126018} {\bibfield  {journal} {\bibinfo  {journal} {Phys. Rev. D}\ }\textbf {\bibinfo {volume} {101}},\ \bibinfo {pages} {126018} (\bibinfo {year} {2020})},\ \Eprint {https://arxiv.org/abs/2004.13140} {arXiv:2004.13140 [hep-th]} \BibitemShut {NoStop}%
\bibitem [{\citenamefont {Elgood}\ and\ \citenamefont {Ortin}(2020)}]{Elgood:2020xwu}%
  \BibitemOpen
  \bibfield  {author} {\bibinfo {author} {\bibfnamefont {Z.}~\bibnamefont {Elgood}}\ and\ \bibinfo {author} {\bibfnamefont {T.}~\bibnamefont {Ortin}},\ }\bibfield  {title} {\bibinfo {title} {{T duality and Wald entropy formula in the Heterotic Superstring effective action at first-order in \ensuremath{\alpha}'}},\ }\href {https://doi.org/10.1007/JHEP10(2020)097} {\bibfield  {journal} {\bibinfo  {journal} {JHEP}\ }\textbf {\bibinfo {volume} {10}},\ \bibinfo {pages} {097}},\ \bibinfo {note} {[Erratum: JHEP 06, 105 (2021)]},\ \Eprint {https://arxiv.org/abs/2005.11272} {arXiv:2005.11272 [hep-th]} \BibitemShut {NoStop}%
\bibitem [{\citenamefont {Ortin}(2021)}]{Ortin:2020xdm}%
  \BibitemOpen
  \bibfield  {author} {\bibinfo {author} {\bibfnamefont {T.}~\bibnamefont {Ortin}},\ }\bibfield  {title} {\bibinfo {title} {{O(n, n) invariance and Wald entropy formula in the Heterotic Superstring effective action at first order in $\alpha'$}},\ }\href {https://doi.org/10.1007/JHEP01(2021)187} {\bibfield  {journal} {\bibinfo  {journal} {JHEP}\ }\textbf {\bibinfo {volume} {01}},\ \bibinfo {pages} {187}},\ \Eprint {https://arxiv.org/abs/2005.14618} {arXiv:2005.14618 [hep-th]} \BibitemShut {NoStop}%
\bibitem [{\citenamefont {Jayaprakash}\ and\ \citenamefont {Liu}(2024)}]{Jayaprakash:2024xlr}%
  \BibitemOpen
  \bibfield  {author} {\bibinfo {author} {\bibfnamefont {S.}~\bibnamefont {Jayaprakash}}\ and\ \bibinfo {author} {\bibfnamefont {J.~T.}\ \bibnamefont {Liu}},\ }\bibfield  {title} {\bibinfo {title} {{Higher derivative heterotic supergravity on a torus and supersymmetry}},\ }\href {https://doi.org/10.1007/JHEP12(2024)076} {\bibfield  {journal} {\bibinfo  {journal} {JHEP}\ }\textbf {\bibinfo {volume} {12}},\ \bibinfo {pages} {076}},\ \Eprint {https://arxiv.org/abs/2406.14600} {arXiv:2406.14600 [hep-th]} \BibitemShut {NoStop}%
\bibitem [{\citenamefont {Kaloper}\ and\ \citenamefont {Meissner}(1997)}]{Kaloper:1997ux}%
  \BibitemOpen
  \bibfield  {author} {\bibinfo {author} {\bibfnamefont {N.}~\bibnamefont {Kaloper}}\ and\ \bibinfo {author} {\bibfnamefont {K.~A.}\ \bibnamefont {Meissner}},\ }\bibfield  {title} {\bibinfo {title} {{Duality beyond the first loop}},\ }\href {https://doi.org/10.1103/PhysRevD.56.7940} {\bibfield  {journal} {\bibinfo  {journal} {Phys. Rev. D}\ }\textbf {\bibinfo {volume} {56}},\ \bibinfo {pages} {7940} (\bibinfo {year} {1997})},\ \Eprint {https://arxiv.org/abs/hep-th/9705193} {arXiv:hep-th/9705193} \BibitemShut {NoStop}%
\bibitem [{\citenamefont {Garousi}(2019)}]{Garousi:2019wgz}%
  \BibitemOpen
  \bibfield  {author} {\bibinfo {author} {\bibfnamefont {M.~R.}\ \bibnamefont {Garousi}},\ }\bibfield  {title} {\bibinfo {title} {{Four-derivative couplings via the $T$-duality invariance constraint}},\ }\href {https://doi.org/10.1103/PhysRevD.99.126005} {\bibfield  {journal} {\bibinfo  {journal} {Phys. Rev. D}\ }\textbf {\bibinfo {volume} {99}},\ \bibinfo {pages} {126005} (\bibinfo {year} {2019})},\ \Eprint {https://arxiv.org/abs/1904.11282} {arXiv:1904.11282 [hep-th]} \BibitemShut {NoStop}%
\bibitem [{\citenamefont {Xia}\ \emph {et~al.}(2026)\citenamefont {Xia}, \citenamefont {Ma}, \citenamefont {Pang},\ and\ \citenamefont {Saskowski}}]{Xia:2025lvn}%
  \BibitemOpen
  \bibfield  {author} {\bibinfo {author} {\bibfnamefont {M.}~\bibnamefont {Xia}}, \bibinfo {author} {\bibfnamefont {L.}~\bibnamefont {Ma}}, \bibinfo {author} {\bibfnamefont {Y.}~\bibnamefont {Pang}},\ and\ \bibinfo {author} {\bibfnamefont {R.~J.}\ \bibnamefont {Saskowski}},\ }\bibfield  {title} {\bibinfo {title} {{Consistent four-derivative heterotic truncations and the Kerr-Sen solution}},\ }\href {https://doi.org/10.1103/h7kw-7cym} {\bibfield  {journal} {\bibinfo  {journal} {Phys. Rev. D}\ }\textbf {\bibinfo {volume} {113}},\ \bibinfo {pages} {046019} (\bibinfo {year} {2026})},\ \Eprint {https://arxiv.org/abs/2509.07069} {arXiv:2509.07069 [hep-th]} \BibitemShut {NoStop}%
\bibitem [{\citenamefont {Hu}\ \emph {et~al.}(2026)\citenamefont {Hu}, \citenamefont {Ma}, \citenamefont {Pang},\ and\ \citenamefont {Saskowski}}]{Hu:2025aji}%
  \BibitemOpen
  \bibfield  {author} {\bibinfo {author} {\bibfnamefont {P.-J.}\ \bibnamefont {Hu}}, \bibinfo {author} {\bibfnamefont {L.}~\bibnamefont {Ma}}, \bibinfo {author} {\bibfnamefont {Y.}~\bibnamefont {Pang}},\ and\ \bibinfo {author} {\bibfnamefont {R.~J.}\ \bibnamefont {Saskowski}},\ }\bibfield  {title} {\bibinfo {title} {{Higher-derivative heterotic Kerr-Sen black holes}},\ }\href {https://doi.org/10.1007/JHEP02(2026)235} {\bibfield  {journal} {\bibinfo  {journal} {JHEP}\ }\textbf {\bibinfo {volume} {02}},\ \bibinfo {pages} {235}},\ \Eprint {https://arxiv.org/abs/2506.20077} {arXiv:2506.20077 [hep-th]} \BibitemShut {NoStop}%
\bibitem [{\citenamefont {Baron}\ \emph {et~al.}(2017)\citenamefont {Baron}, \citenamefont {Fernandez-Melgarejo}, \citenamefont {Marques},\ and\ \citenamefont {Nunez}}]{Baron:2017dvb}%
  \BibitemOpen
  \bibfield  {author} {\bibinfo {author} {\bibfnamefont {W.~H.}\ \bibnamefont {Baron}}, \bibinfo {author} {\bibfnamefont {J.~J.}\ \bibnamefont {Fernandez-Melgarejo}}, \bibinfo {author} {\bibfnamefont {D.}~\bibnamefont {Marques}},\ and\ \bibinfo {author} {\bibfnamefont {C.}~\bibnamefont {Nunez}},\ }\bibfield  {title} {\bibinfo {title} {{The Odd story of {\ensuremath{\alpha}}'-corrections}},\ }\href {https://doi.org/10.1007/JHEP04(2017)078} {\bibfield  {journal} {\bibinfo  {journal} {JHEP}\ }\textbf {\bibinfo {volume} {04}},\ \bibinfo {pages} {078}},\ \Eprint {https://arxiv.org/abs/1702.05489} {arXiv:1702.05489 [hep-th]} \BibitemShut {NoStop}%
\bibitem [{\citenamefont {Veneziano}(1991)}]{Veneziano:1991ek}%
  \BibitemOpen
  \bibfield  {author} {\bibinfo {author} {\bibfnamefont {G.}~\bibnamefont {Veneziano}},\ }\bibfield  {title} {\bibinfo {title} {{Scale factor duality for classical and quantum strings}},\ }\href {https://doi.org/10.1016/0370-2693(91)90055-U} {\bibfield  {journal} {\bibinfo  {journal} {Phys. Lett. B}\ }\textbf {\bibinfo {volume} {265}},\ \bibinfo {pages} {287} (\bibinfo {year} {1991})}\BibitemShut {NoStop}%
\bibitem [{\citenamefont {Meissner}\ and\ \citenamefont {Veneziano}(1991)}]{Meissner:1991zj}%
  \BibitemOpen
  \bibfield  {author} {\bibinfo {author} {\bibfnamefont {K.~A.}\ \bibnamefont {Meissner}}\ and\ \bibinfo {author} {\bibfnamefont {G.}~\bibnamefont {Veneziano}},\ }\bibfield  {title} {\bibinfo {title} {{Symmetries of cosmological superstring vacua}},\ }\href {https://doi.org/10.1016/0370-2693(91)90520-Z} {\bibfield  {journal} {\bibinfo  {journal} {Phys. Lett. B}\ }\textbf {\bibinfo {volume} {267}},\ \bibinfo {pages} {33} (\bibinfo {year} {1991})}\BibitemShut {NoStop}%
\bibitem [{\citenamefont {Sen}(1992{\natexlab{a}})}]{Sen:1991cn}%
  \BibitemOpen
  \bibfield  {author} {\bibinfo {author} {\bibfnamefont {A.}~\bibnamefont {Sen}},\ }\bibfield  {title} {\bibinfo {title} {{Twisted black p-brane solutions in string theory}},\ }\href {https://doi.org/10.1016/0370-2693(92)90300-S} {\bibfield  {journal} {\bibinfo  {journal} {Phys. Lett. B}\ }\textbf {\bibinfo {volume} {274}},\ \bibinfo {pages} {34} (\bibinfo {year} {1992}{\natexlab{a}})},\ \Eprint {https://arxiv.org/abs/hep-th/9108011} {arXiv:hep-th/9108011} \BibitemShut {NoStop}%
\bibitem [{\citenamefont {Gasperini}\ \emph {et~al.}(1991)\citenamefont {Gasperini}, \citenamefont {Maharana},\ and\ \citenamefont {Veneziano}}]{Gasperini:1991qy}%
  \BibitemOpen
  \bibfield  {author} {\bibinfo {author} {\bibfnamefont {M.}~\bibnamefont {Gasperini}}, \bibinfo {author} {\bibfnamefont {J.}~\bibnamefont {Maharana}},\ and\ \bibinfo {author} {\bibfnamefont {G.}~\bibnamefont {Veneziano}},\ }\bibfield  {title} {\bibinfo {title} {{From trivial to nontrivial conformal string backgrounds via O(d,d) transformations}},\ }\href {https://doi.org/10.1016/0370-2693(91)91831-F} {\bibfield  {journal} {\bibinfo  {journal} {Phys. Lett. B}\ }\textbf {\bibinfo {volume} {272}},\ \bibinfo {pages} {277} (\bibinfo {year} {1991})}\BibitemShut {NoStop}%
\bibitem [{\citenamefont {Sen}(1992{\natexlab{b}})}]{Sen:1992ua}%
  \BibitemOpen
  \bibfield  {author} {\bibinfo {author} {\bibfnamefont {A.}~\bibnamefont {Sen}},\ }\bibfield  {title} {\bibinfo {title} {{Rotating charged black hole solution in heterotic string theory}},\ }\href {https://doi.org/10.1103/PhysRevLett.69.1006} {\bibfield  {journal} {\bibinfo  {journal} {Phys. Rev. Lett.}\ }\textbf {\bibinfo {volume} {69}},\ \bibinfo {pages} {1006} (\bibinfo {year} {1992}{\natexlab{b}})},\ \Eprint {https://arxiv.org/abs/hep-th/9204046} {arXiv:hep-th/9204046} \BibitemShut {NoStop}%
\bibitem [{\citenamefont {Cvetic}\ and\ \citenamefont {Youm}(1995)}]{Cvetic:1995sz}%
  \BibitemOpen
  \bibfield  {author} {\bibinfo {author} {\bibfnamefont {M.}~\bibnamefont {Cvetic}}\ and\ \bibinfo {author} {\bibfnamefont {D.}~\bibnamefont {Youm}},\ }\bibfield  {title} {\bibinfo {title} {{All the four-dimensional static, spherically symmetric solutions of Abelian Kaluza-Klein theory}},\ }\href {https://doi.org/10.1103/PhysRevLett.75.4165} {\bibfield  {journal} {\bibinfo  {journal} {Phys. Rev. Lett.}\ }\textbf {\bibinfo {volume} {75}},\ \bibinfo {pages} {4165} (\bibinfo {year} {1995})},\ \Eprint {https://arxiv.org/abs/hep-th/9503082} {arXiv:hep-th/9503082} \BibitemShut {NoStop}%
\bibitem [{\citenamefont {Cvetic}\ and\ \citenamefont {Youm}(1996{\natexlab{a}})}]{Cvetic:1995kv}%
  \BibitemOpen
  \bibfield  {author} {\bibinfo {author} {\bibfnamefont {M.}~\bibnamefont {Cvetic}}\ and\ \bibinfo {author} {\bibfnamefont {D.}~\bibnamefont {Youm}},\ }\bibfield  {title} {\bibinfo {title} {{All the static spherically symmetric black holes of heterotic string on a six torus}},\ }\href {https://doi.org/10.1016/0550-3213(96)00219-2} {\bibfield  {journal} {\bibinfo  {journal} {Nucl. Phys. B}\ }\textbf {\bibinfo {volume} {472}},\ \bibinfo {pages} {249} (\bibinfo {year} {1996}{\natexlab{a}})},\ \Eprint {https://arxiv.org/abs/hep-th/9512127} {arXiv:hep-th/9512127} \BibitemShut {NoStop}%
\bibitem [{\citenamefont {Cvetic}\ and\ \citenamefont {Youm}(1996{\natexlab{b}})}]{Cvetic:1996xz}%
  \BibitemOpen
  \bibfield  {author} {\bibinfo {author} {\bibfnamefont {M.}~\bibnamefont {Cvetic}}\ and\ \bibinfo {author} {\bibfnamefont {D.}~\bibnamefont {Youm}},\ }\bibfield  {title} {\bibinfo {title} {{General rotating five-dimensional black holes of toroidally compactified heterotic string}},\ }\href {https://doi.org/10.1016/0550-3213(96)00355-0} {\bibfield  {journal} {\bibinfo  {journal} {Nucl. Phys. B}\ }\textbf {\bibinfo {volume} {476}},\ \bibinfo {pages} {118} (\bibinfo {year} {1996}{\natexlab{b}})},\ \Eprint {https://arxiv.org/abs/hep-th/9603100} {arXiv:hep-th/9603100} \BibitemShut {NoStop}%
\bibitem [{\citenamefont {Font}\ \emph {et~al.}(1990)\citenamefont {Font}, \citenamefont {Ibanez}, \citenamefont {Lust},\ and\ \citenamefont {Quevedo}}]{Font:1990gx}%
  \BibitemOpen
  \bibfield  {author} {\bibinfo {author} {\bibfnamefont {A.}~\bibnamefont {Font}}, \bibinfo {author} {\bibfnamefont {L.~E.}\ \bibnamefont {Ibanez}}, \bibinfo {author} {\bibfnamefont {D.}~\bibnamefont {Lust}},\ and\ \bibinfo {author} {\bibfnamefont {F.}~\bibnamefont {Quevedo}},\ }\bibfield  {title} {\bibinfo {title} {{Strong - weak coupling duality and nonperturbative effects in string theory}},\ }\href {https://doi.org/10.1016/0370-2693(90)90523-9} {\bibfield  {journal} {\bibinfo  {journal} {Phys. Lett. B}\ }\textbf {\bibinfo {volume} {249}},\ \bibinfo {pages} {35} (\bibinfo {year} {1990})}\BibitemShut {NoStop}%
\bibitem [{\citenamefont {Duff}(1995)}]{Duff:1994zt}%
  \BibitemOpen
  \bibfield  {author} {\bibinfo {author} {\bibfnamefont {M.~J.}\ \bibnamefont {Duff}},\ }\bibfield  {title} {\bibinfo {title} {{Strong / weak coupling duality from the dual string}},\ }\href {https://doi.org/10.1016/S0550-3213(95)00070-4} {\bibfield  {journal} {\bibinfo  {journal} {Nucl. Phys. B}\ }\textbf {\bibinfo {volume} {442}},\ \bibinfo {pages} {47} (\bibinfo {year} {1995})},\ \Eprint {https://arxiv.org/abs/hep-th/9501030} {arXiv:hep-th/9501030} \BibitemShut {NoStop}%
\bibitem [{\citenamefont {Sen}(1994)}]{Sen:1994fa}%
  \BibitemOpen
  \bibfield  {author} {\bibinfo {author} {\bibfnamefont {A.}~\bibnamefont {Sen}},\ }\bibfield  {title} {\bibinfo {title} {{Strong - weak coupling duality in four-dimensional string theory}},\ }\href {https://doi.org/10.1142/S0217751X94001497} {\bibfield  {journal} {\bibinfo  {journal} {Int. J. Mod. Phys. A}\ }\textbf {\bibinfo {volume} {9}},\ \bibinfo {pages} {3707} (\bibinfo {year} {1994})},\ \Eprint {https://arxiv.org/abs/hep-th/9402002} {arXiv:hep-th/9402002} \BibitemShut {NoStop}%
\bibitem [{\citenamefont {Sen}(1995)}]{Sen:1994wr}%
  \BibitemOpen
  \bibfield  {author} {\bibinfo {author} {\bibfnamefont {A.}~\bibnamefont {Sen}},\ }\bibfield  {title} {\bibinfo {title} {{Strong - weak coupling duality in three-dimensional string theory}},\ }\href {https://doi.org/10.1016/0550-3213(94)00461-M} {\bibfield  {journal} {\bibinfo  {journal} {Nucl. Phys. B}\ }\textbf {\bibinfo {volume} {434}},\ \bibinfo {pages} {179} (\bibinfo {year} {1995})},\ \Eprint {https://arxiv.org/abs/hep-th/9408083} {arXiv:hep-th/9408083} \BibitemShut {NoStop}%
\bibitem [{\citenamefont {Cremmer}\ \emph {et~al.}(1998{\natexlab{a}})\citenamefont {Cremmer}, \citenamefont {Julia}, \citenamefont {Lu},\ and\ \citenamefont {Pope}}]{Cremmer:1997ct}%
  \BibitemOpen
  \bibfield  {author} {\bibinfo {author} {\bibfnamefont {E.}~\bibnamefont {Cremmer}}, \bibinfo {author} {\bibfnamefont {B.}~\bibnamefont {Julia}}, \bibinfo {author} {\bibfnamefont {H.}~\bibnamefont {Lu}},\ and\ \bibinfo {author} {\bibfnamefont {C.~N.}\ \bibnamefont {Pope}},\ }\bibfield  {title} {\bibinfo {title} {{Dualization of dualities. 1.}},\ }\href {https://doi.org/10.1016/S0550-3213(98)00136-9} {\bibfield  {journal} {\bibinfo  {journal} {Nucl. Phys. B}\ }\textbf {\bibinfo {volume} {523}},\ \bibinfo {pages} {73} (\bibinfo {year} {1998}{\natexlab{a}})},\ \Eprint {https://arxiv.org/abs/hep-th/9710119} {arXiv:hep-th/9710119} \BibitemShut {NoStop}%
\bibitem [{\citenamefont {Cremmer}\ \emph {et~al.}(1998{\natexlab{b}})\citenamefont {Cremmer}, \citenamefont {Julia}, \citenamefont {Lu},\ and\ \citenamefont {Pope}}]{Cremmer:1998px}%
  \BibitemOpen
  \bibfield  {author} {\bibinfo {author} {\bibfnamefont {E.}~\bibnamefont {Cremmer}}, \bibinfo {author} {\bibfnamefont {B.}~\bibnamefont {Julia}}, \bibinfo {author} {\bibfnamefont {H.}~\bibnamefont {Lu}},\ and\ \bibinfo {author} {\bibfnamefont {C.~N.}\ \bibnamefont {Pope}},\ }\bibfield  {title} {\bibinfo {title} {{Dualization of dualities. 2. Twisted self-duality of doubled fields, and superdualities}},\ }\href {https://doi.org/10.1016/S0550-3213(98)00552-5} {\bibfield  {journal} {\bibinfo  {journal} {Nucl. Phys. B}\ }\textbf {\bibinfo {volume} {535}},\ \bibinfo {pages} {242} (\bibinfo {year} {1998}{\natexlab{b}})},\ \Eprint {https://arxiv.org/abs/hep-th/9806106} {arXiv:hep-th/9806106} \BibitemShut {NoStop}%
\bibitem [{\citenamefont {Cecotti}\ \emph {et~al.}(1989)\citenamefont {Cecotti}, \citenamefont {Ferrara},\ and\ \citenamefont {Girardello}}]{Cecotti:1988qn}%
  \BibitemOpen
  \bibfield  {author} {\bibinfo {author} {\bibfnamefont {S.}~\bibnamefont {Cecotti}}, \bibinfo {author} {\bibfnamefont {S.}~\bibnamefont {Ferrara}},\ and\ \bibinfo {author} {\bibfnamefont {L.}~\bibnamefont {Girardello}},\ }\bibfield  {title} {\bibinfo {title} {{Geometry of Type II Superstrings and the Moduli of Superconformal Field Theories}},\ }\href {https://doi.org/10.1142/S0217751X89000972} {\bibfield  {journal} {\bibinfo  {journal} {Int. J. Mod. Phys. A}\ }\textbf {\bibinfo {volume} {4}},\ \bibinfo {pages} {2475} (\bibinfo {year} {1989})}\BibitemShut {NoStop}%
\bibitem [{\citenamefont {Mizoguchi}\ and\ \citenamefont {Ohta}(1998)}]{Mizoguchi:1998wv}%
  \BibitemOpen
  \bibfield  {author} {\bibinfo {author} {\bibfnamefont {S.}~\bibnamefont {Mizoguchi}}\ and\ \bibinfo {author} {\bibfnamefont {N.}~\bibnamefont {Ohta}},\ }\bibfield  {title} {\bibinfo {title} {{More on the similarity between D = 5 simple supergravity and M theory}},\ }\href {https://doi.org/10.1016/S0370-2693(98)01122-8} {\bibfield  {journal} {\bibinfo  {journal} {Phys. Lett. B}\ }\textbf {\bibinfo {volume} {441}},\ \bibinfo {pages} {123} (\bibinfo {year} {1998})},\ \Eprint {https://arxiv.org/abs/hep-th/9807111} {arXiv:hep-th/9807111} \BibitemShut {NoStop}%
\bibitem [{\citenamefont {Cremmer}\ \emph {et~al.}(1999)\citenamefont {Cremmer}, \citenamefont {Julia}, \citenamefont {Lu},\ and\ \citenamefont {Pope}}]{Cremmer:1999du}%
  \BibitemOpen
  \bibfield  {author} {\bibinfo {author} {\bibfnamefont {E.}~\bibnamefont {Cremmer}}, \bibinfo {author} {\bibfnamefont {B.}~\bibnamefont {Julia}}, \bibinfo {author} {\bibfnamefont {H.}~\bibnamefont {Lu}},\ and\ \bibinfo {author} {\bibfnamefont {C.~N.}\ \bibnamefont {Pope}},\ }\bibfield  {title} {\bibinfo {title} {{Higher dimensional origin of D = 3 coset symmetries}},\ }\Eprint {https://arxiv.org/abs/hep-th/9909099} {arXiv:hep-th/9909099}  (\bibinfo {year} {1999})\BibitemShut {NoStop}%
\bibitem [{\citenamefont {Bouchareb}\ \emph {et~al.}(2007)\citenamefont {Bouchareb}, \citenamefont {Clement}, \citenamefont {Chen}, \citenamefont {Gal'tsov}, \citenamefont {Scherbluk},\ and\ \citenamefont {Wolf}}]{Bouchareb:2007ax}%
  \BibitemOpen
  \bibfield  {author} {\bibinfo {author} {\bibfnamefont {A.}~\bibnamefont {Bouchareb}}, \bibinfo {author} {\bibfnamefont {G.}~\bibnamefont {Clement}}, \bibinfo {author} {\bibfnamefont {C.-M.}\ \bibnamefont {Chen}}, \bibinfo {author} {\bibfnamefont {D.~V.}\ \bibnamefont {Gal'tsov}}, \bibinfo {author} {\bibfnamefont {N.~G.}\ \bibnamefont {Scherbluk}},\ and\ \bibinfo {author} {\bibfnamefont {T.}~\bibnamefont {Wolf}},\ }\bibfield  {title} {\bibinfo {title} {{G(2) generating technique for minimal D=5 supergravity and black rings}},\ }\href {https://doi.org/10.1103/PhysRevD.76.104032} {\bibfield  {journal} {\bibinfo  {journal} {Phys. Rev. D}\ }\textbf {\bibinfo {volume} {76}},\ \bibinfo {pages} {104032} (\bibinfo {year} {2007})},\ \bibinfo {note} {[Erratum: Phys.Rev.D 78, 029901 (2008)]},\ \Eprint {https://arxiv.org/abs/0708.2361} {arXiv:0708.2361 [hep-th]} \BibitemShut {NoStop}%
\bibitem [{\citenamefont {Tomizawa}\ \emph {et~al.}(2009)\citenamefont {Tomizawa}, \citenamefont {Yasui},\ and\ \citenamefont {Morisawa}}]{Tomizawa:2008qr}%
  \BibitemOpen
  \bibfield  {author} {\bibinfo {author} {\bibfnamefont {S.}~\bibnamefont {Tomizawa}}, \bibinfo {author} {\bibfnamefont {Y.}~\bibnamefont {Yasui}},\ and\ \bibinfo {author} {\bibfnamefont {Y.}~\bibnamefont {Morisawa}},\ }\bibfield  {title} {\bibinfo {title} {{Charged Rotating Kaluza-Klein Black Holes Generated by G2(2) Transformation}},\ }\href {https://doi.org/10.1088/0264-9381/26/14/145006} {\bibfield  {journal} {\bibinfo  {journal} {Class. Quant. Grav.}\ }\textbf {\bibinfo {volume} {26}},\ \bibinfo {pages} {145006} (\bibinfo {year} {2009})},\ \Eprint {https://arxiv.org/abs/0809.2001} {arXiv:0809.2001 [hep-th]} \BibitemShut {NoStop}%
\bibitem [{\citenamefont {Compere}\ \emph {et~al.}(2009)\citenamefont {Compere}, \citenamefont {de~Buyl}, \citenamefont {Jamsin},\ and\ \citenamefont {Virmani}}]{Compere:2009zh}%
  \BibitemOpen
  \bibfield  {author} {\bibinfo {author} {\bibfnamefont {G.}~\bibnamefont {Compere}}, \bibinfo {author} {\bibfnamefont {S.}~\bibnamefont {de~Buyl}}, \bibinfo {author} {\bibfnamefont {E.}~\bibnamefont {Jamsin}},\ and\ \bibinfo {author} {\bibfnamefont {A.}~\bibnamefont {Virmani}},\ }\bibfield  {title} {\bibinfo {title} {{$G_2$ Dualities in $D=5$ Supergravity and Black Strings}},\ }\href {https://doi.org/10.1088/0264-9381/26/12/125016} {\bibfield  {journal} {\bibinfo  {journal} {Class. Quant. Grav.}\ }\textbf {\bibinfo {volume} {26}},\ \bibinfo {pages} {125016} (\bibinfo {year} {2009})},\ \Eprint {https://arxiv.org/abs/0903.1645} {arXiv:0903.1645 [hep-th]} \BibitemShut {NoStop}%
\bibitem [{\citenamefont {Suzuki}\ and\ \citenamefont {Tomizawa}(2024{\natexlab{a}})}]{Suzuki:2024coe}%
  \BibitemOpen
  \bibfield  {author} {\bibinfo {author} {\bibfnamefont {R.}~\bibnamefont {Suzuki}}\ and\ \bibinfo {author} {\bibfnamefont {S.}~\bibnamefont {Tomizawa}},\ }\bibfield  {title} {\bibinfo {title} {{New construction of a charged dipole black ring by the Harrison transformation}},\ }\href {https://doi.org/10.1103/PhysRevD.109.084020} {\bibfield  {journal} {\bibinfo  {journal} {Phys. Rev. D}\ }\textbf {\bibinfo {volume} {109}},\ \bibinfo {pages} {084020} (\bibinfo {year} {2024}{\natexlab{a}})},\ \Eprint {https://arxiv.org/abs/2402.07589} {arXiv:2402.07589 [hep-th]} \BibitemShut {NoStop}%
\bibitem [{\citenamefont {Suzuki}\ and\ \citenamefont {Tomizawa}(2024{\natexlab{b}})}]{Suzuki:2024phv}%
  \BibitemOpen
  \bibfield  {author} {\bibinfo {author} {\bibfnamefont {R.}~\bibnamefont {Suzuki}}\ and\ \bibinfo {author} {\bibfnamefont {S.}~\bibnamefont {Tomizawa}},\ }\bibfield  {title} {\bibinfo {title} {{Solution generation of a capped black hole}},\ }\href {https://doi.org/10.1103/PhysRevD.110.024026} {\bibfield  {journal} {\bibinfo  {journal} {Phys. Rev. D}\ }\textbf {\bibinfo {volume} {110}},\ \bibinfo {pages} {024026} (\bibinfo {year} {2024}{\natexlab{b}})},\ \Eprint {https://arxiv.org/abs/2403.17796} {arXiv:2403.17796 [hep-th]} \BibitemShut {NoStop}%
\bibitem [{\citenamefont {Suzuki}\ and\ \citenamefont {Tomizawa}(2024{\natexlab{c}})}]{Suzuki:2024vzq}%
  \BibitemOpen
  \bibfield  {author} {\bibinfo {author} {\bibfnamefont {R.}~\bibnamefont {Suzuki}}\ and\ \bibinfo {author} {\bibfnamefont {S.}~\bibnamefont {Tomizawa}},\ }\bibfield  {title} {\bibinfo {title} {{New black ring with all independent conserved charges in five-dimensional minimal supergravity}},\ }\href {https://doi.org/10.1103/PhysRevD.110.124041} {\bibfield  {journal} {\bibinfo  {journal} {Phys. Rev. D}\ }\textbf {\bibinfo {volume} {110}},\ \bibinfo {pages} {124041} (\bibinfo {year} {2024}{\natexlab{c}})},\ \Eprint {https://arxiv.org/abs/2407.18142} {arXiv:2407.18142 [hep-th]} \BibitemShut {NoStop}%
\bibitem [{\citenamefont {Chong}\ \emph {et~al.}(2005)\citenamefont {Chong}, \citenamefont {Cvetic}, \citenamefont {Lu},\ and\ \citenamefont {Pope}}]{Chong:2004na}%
  \BibitemOpen
  \bibfield  {author} {\bibinfo {author} {\bibfnamefont {Z.~W.}\ \bibnamefont {Chong}}, \bibinfo {author} {\bibfnamefont {M.}~\bibnamefont {Cvetic}}, \bibinfo {author} {\bibfnamefont {H.}~\bibnamefont {Lu}},\ and\ \bibinfo {author} {\bibfnamefont {C.~N.}\ \bibnamefont {Pope}},\ }\bibfield  {title} {\bibinfo {title} {{Charged rotating black holes in four-dimensional gauged and ungauged supergravities}},\ }\href {https://doi.org/10.1016/j.nuclphysb.2005.03.034} {\bibfield  {journal} {\bibinfo  {journal} {Nucl. Phys. B}\ }\textbf {\bibinfo {volume} {717}},\ \bibinfo {pages} {246} (\bibinfo {year} {2005})},\ \Eprint {https://arxiv.org/abs/hep-th/0411045} {arXiv:hep-th/0411045} \BibitemShut {NoStop}%
\bibitem [{\citenamefont {Gal'tsov}\ and\ \citenamefont {Scherbluk}(2008)}]{Galtsov:2008bmt}%
  \BibitemOpen
  \bibfield  {author} {\bibinfo {author} {\bibfnamefont {D.~V.}\ \bibnamefont {Gal'tsov}}\ and\ \bibinfo {author} {\bibfnamefont {N.~G.}\ \bibnamefont {Scherbluk}},\ }\bibfield  {title} {\bibinfo {title} {{Generating technique for U(1)**3 5D supergravity}},\ }\href {https://doi.org/10.1103/PhysRevD.78.064033} {\bibfield  {journal} {\bibinfo  {journal} {Phys. Rev. D}\ }\textbf {\bibinfo {volume} {78}},\ \bibinfo {pages} {064033} (\bibinfo {year} {2008})},\ \Eprint {https://arxiv.org/abs/0805.3924} {arXiv:0805.3924 [hep-th]} \BibitemShut {NoStop}%
\bibitem [{\citenamefont {Gal'tsov}\ and\ \citenamefont {Scherbluk}(2009)}]{Galtsov:2008jjb}%
  \BibitemOpen
  \bibfield  {author} {\bibinfo {author} {\bibfnamefont {D.~V.}\ \bibnamefont {Gal'tsov}}\ and\ \bibinfo {author} {\bibfnamefont {N.~G.}\ \bibnamefont {Scherbluk}},\ }\bibfield  {title} {\bibinfo {title} {{Improved generating technique for D=5 supergravities and squashed Kaluza-Klein Black Holes}},\ }\href {https://doi.org/10.1103/PhysRevD.79.064020} {\bibfield  {journal} {\bibinfo  {journal} {Phys. Rev. D}\ }\textbf {\bibinfo {volume} {79}},\ \bibinfo {pages} {064020} (\bibinfo {year} {2009})},\ \Eprint {https://arxiv.org/abs/0812.2336} {arXiv:0812.2336 [hep-th]} \BibitemShut {NoStop}%
\bibitem [{\citenamefont {Lambert}\ and\ \citenamefont {West}(2006)}]{Lambert:2006he}%
  \BibitemOpen
  \bibfield  {author} {\bibinfo {author} {\bibfnamefont {N.}~\bibnamefont {Lambert}}\ and\ \bibinfo {author} {\bibfnamefont {P.~C.}\ \bibnamefont {West}},\ }\bibfield  {title} {\bibinfo {title} {{Enhanced Coset Symmetries and Higher Derivative Corrections}},\ }\href {https://doi.org/10.1103/PhysRevD.74.065002} {\bibfield  {journal} {\bibinfo  {journal} {Phys. Rev. D}\ }\textbf {\bibinfo {volume} {74}},\ \bibinfo {pages} {065002} (\bibinfo {year} {2006})},\ \Eprint {https://arxiv.org/abs/hep-th/0603255} {arXiv:hep-th/0603255} \BibitemShut {NoStop}%
\bibitem [{\citenamefont {Bao}\ \emph {et~al.}(2008)\citenamefont {Bao}, \citenamefont {Cederwall},\ and\ \citenamefont {Nilsson}}]{Bao:2007er}%
  \BibitemOpen
  \bibfield  {author} {\bibinfo {author} {\bibfnamefont {L.}~\bibnamefont {Bao}}, \bibinfo {author} {\bibfnamefont {M.}~\bibnamefont {Cederwall}},\ and\ \bibinfo {author} {\bibfnamefont {B.~E.~W.}\ \bibnamefont {Nilsson}},\ }\bibfield  {title} {\bibinfo {title} {{Aspects of higher curvature terms and U-duality}},\ }\href {https://doi.org/10.1088/0264-9381/25/9/095001} {\bibfield  {journal} {\bibinfo  {journal} {Class. Quant. Grav.}\ }\textbf {\bibinfo {volume} {25}},\ \bibinfo {pages} {095001} (\bibinfo {year} {2008})},\ \Eprint {https://arxiv.org/abs/0706.1183} {arXiv:0706.1183 [hep-th]} \BibitemShut {NoStop}%
\bibitem [{\citenamefont {Eloy}\ \emph {et~al.}(2023)\citenamefont {Eloy}, \citenamefont {Hohm},\ and\ \citenamefont {Samtleben}}]{Eloy:2022vsq}%
  \BibitemOpen
  \bibfield  {author} {\bibinfo {author} {\bibfnamefont {C.}~\bibnamefont {Eloy}}, \bibinfo {author} {\bibfnamefont {O.}~\bibnamefont {Hohm}},\ and\ \bibinfo {author} {\bibfnamefont {H.}~\bibnamefont {Samtleben}},\ }\bibfield  {title} {\bibinfo {title} {{U duality and {\ensuremath{\alpha}}' corrections in three dimensions}},\ }\href {https://doi.org/10.1103/PhysRevD.108.026015} {\bibfield  {journal} {\bibinfo  {journal} {Phys. Rev. D}\ }\textbf {\bibinfo {volume} {108}},\ \bibinfo {pages} {026015} (\bibinfo {year} {2023})},\ \Eprint {https://arxiv.org/abs/2211.16358} {arXiv:2211.16358 [hep-th]} \BibitemShut {NoStop}%
\bibitem [{\citenamefont {Pang}\ and\ \citenamefont {Saskowski}(2026)}]{Pang:2026urr}%
  \BibitemOpen
  \bibfield  {author} {\bibinfo {author} {\bibfnamefont {Y.}~\bibnamefont {Pang}}\ and\ \bibinfo {author} {\bibfnamefont {R.~J.}\ \bibnamefont {Saskowski}},\ }\bibfield  {title} {\bibinfo {title} {{Symmetries of non-maximal supergravities with higher-derivative corrections}},\ }\Eprint {https://arxiv.org/abs/2603.21459} {arXiv:2603.21459 [hep-th]}  (\bibinfo {year} {2026})\BibitemShut {NoStop}%
\bibitem [{\citenamefont {Sen}(2020)}]{Sen:2020cef}%
  \BibitemOpen
  \bibfield  {author} {\bibinfo {author} {\bibfnamefont {A.}~\bibnamefont {Sen}},\ }\bibfield  {title} {\bibinfo {title} {{D-instanton Perturbation Theory}},\ }\href {https://doi.org/10.1007/JHEP08(2020)075} {\bibfield  {journal} {\bibinfo  {journal} {JHEP}\ }\textbf {\bibinfo {volume} {08}},\ \bibinfo {pages} {075}},\ \Eprint {https://arxiv.org/abs/2002.04043} {arXiv:2002.04043 [hep-th]} \BibitemShut {NoStop}%
\end{thebibliography}%
\let\addcontentsline\oldaddcontentsline

\end{document}